\documentclass[
reprint,
amsmath,amssymb,
aps,
prl,
]{revtex4-2}

\usepackage{float}
\usepackage{mathtools}
\usepackage{graphicx}
\usepackage{dcolumn}
\usepackage{bm}
\usepackage{upgreek}
\usepackage{lipsum, babel, xcolor}
\usepackage{amsmath, amssymb}
\usepackage{comment}
\usepackage{natbib}
\usepackage[normalem]{ulem}
\usepackage{physics}
\usepackage{hyperref}
\hypersetup{colorlinks=true,citecolor=blue}

\usepackage[T1]{fontenc}
\usepackage{mathptmx}

\begin{document}


\title{Stable thin-film lithium tantalate modulators operating at high temperature for uncooled operation}

\author{Ayed Al Sayem\textsuperscript{$\dagger$}, Shiekh Zia Uddin\textsuperscript{$\dagger$}, Ting-Chen Hu, Alaric Tate, Mark Cappuzzo, Rose Kopf, Mark Earnshaw}

\affiliation{%
  Nokia Bell Labs, Murray Hill, NJ, USA 
}%
\date{\today}

\begin{abstract}
We demonstrate stable operation of a thin-film lithium tantalate (TFLT) modulator at very high operating temperatures. We show that the electro-optic modulation and bandwidth of the TFLT modulators are not affected by high-temperature operation, and both waveguide and resonant modulators are DC-bias stable even at $\sim\mathrm{120^{o}C}$. At higher temperature, we even observe $\mathrm{\sim10\%}$ reduction of the $\mathrm{V_{\pi}}$ of the modulator. Our results position TFLT modulators as a strong candidate for uncooled operation in co-packaged optics.
\end{abstract}

\maketitle

\section{Introduction}















Co-packaged optics (CPO) has recently emerged as a promising route toward scale-up and scale-out architectures, driven by the ever-growing bandwidth and energy-efficiency demands of artificial intelligence (AI) and data-centric computing systems \cite{Maniotis2024CPO,shekhar2024roadmapping}. By integrating optical transmitters and receivers in close proximity to compute silicon electronics, CPO minimizes electrical trace lengths, thereby boosting bandwidth density and energy efficiency \cite{Minkenberg2021CPO,Tan2023CPO}. However, this tight integration exposes photonic components to elevated local temperatures and significant thermal gradients generated by the nearby silicon compute electronics. Therefore, thermal management and temperature-sensitivity of devices are first-order concerns \cite{Tan2023CPO}. For a practical CPO system, one requires a compact photonic platform that can maintain modulation efficiency, extinction, and high-speed performance across a wide operating range of temperatures with minimal additional power dissipation from thermal control overhead. From a device perspective, temperature-resilient modulators and transmitters are attractive because they reduce or eliminate the need for active cooling and simplify stabilization loops. While silicon microring modulators are currently the frontrunner choice for CPO integration, SiGe-based modulators are gaining traction as a complementary technology \cite{Fujikata2023GeSi85C,yuan20245}. Silicon’s inherently high thermal sensitivity leads to resonance shifts in silicon-based MZMs and microring modulators that complicate dense integration \cite{yang2026thermal}. SiGe electro-absorption modulators (EAMs) offer advantages such as a compact footprint and stable high-temperature operation; however, they are hampered by significant optical loss, a restricted wavelength range, and poor extinction ratios \cite{Fujikata2023GeSi85C,steckler2025monolithic}. The common denominator for both SiGe and silicon based solutions is the need for active thermo-optic (TO) bias control and stabilization. This overhead remains a major barrier to scaling, as it complicates the stabilization necessary for dense, co-packaged systems.

Thin-film Lithium Niobate (TFLN) electro-optic modulators developed in the last decade are now one of the primary candidates for next-generation high-speed optical interconnects \cite{zhu2021integrated,wang2018integrated,kharel2021breaking}. Unfortunately, TFLN suffers strongly from the photo-refractive (PR) effect \cite{xu2021mitigating, Ahmed2025Universal}, which limits the optical power handling, and leads to the DC bias instability \cite{Zhao2025SPIE_DCdift_TFLN,Holzgrafe2024OE_EORelaxation,Powell2024OE_TFLT_DCStable,Celik2024BiasDrift,ren2025photorefractive}. To circumvent DC bias instability, instead of EO bias tuning, thermo-optic (TO) tuning can be used to set the EO modulators at the quadrature point \cite{xu2020high}. But due to the weak thermo-optic (TO) effect in TFLN, resistive heaters require hundreds of mWs of power per modulator \cite{xu2020high}, which is impractical for large-scale, dense photonic circuits where power consumption and excess heat dissipation are major concerns. Furthermore the strong PR effect in TFLN makes resonators optically unstable, especially when operating at higher power levels \cite{xu2021mitigating}, making resonant-based compact modulators extremely challenging in TFLN. Along with thermal-resilience, compactness is another major requirement for CPO because package and shoreline density are central constraints \cite{sturm2025c2po}, and so compact, stable resonant modulator designs are of great interest. We recently showed that Lithium Tantalate (LT) is a highly stable material platform in terms of optical power handling. Due to its low PR effect, LT is $\mathrm{10^{9}}$ times more stabile than LN with oxide cladding, which is a prerequisite for proper velocity matching for high bandwidth operation \cite{kharel2021breaking}. Until now, there have been no experimental demonstrations of TFLT modulators operating at higher temperatures. In this article, we experimentally show the performance of TFLT modulators operating at high temperatures up to $\mathrm{120\,^{o}C}$. We show that the DC bias is stable even at $\mathrm{120\,^{o}C}$ at the quadrature point with a $\sim\mathrm{10\,\%}$ improvement of the $\mathrm{V_{\pi}}$ of the modulator. We also show that the electro-optic bandwidth of a 7\,mm long MZI modulator is beyond 50\,GHz without any noticeable effect when operated at higher temperature. Finally, we show that DC bias tuning works for the resonant modulator on the TFLT platform in a stable manner at elevated temperature.

\begin{figure*}[ht]
    \centering
    \includegraphics[width = 0.9\textwidth]{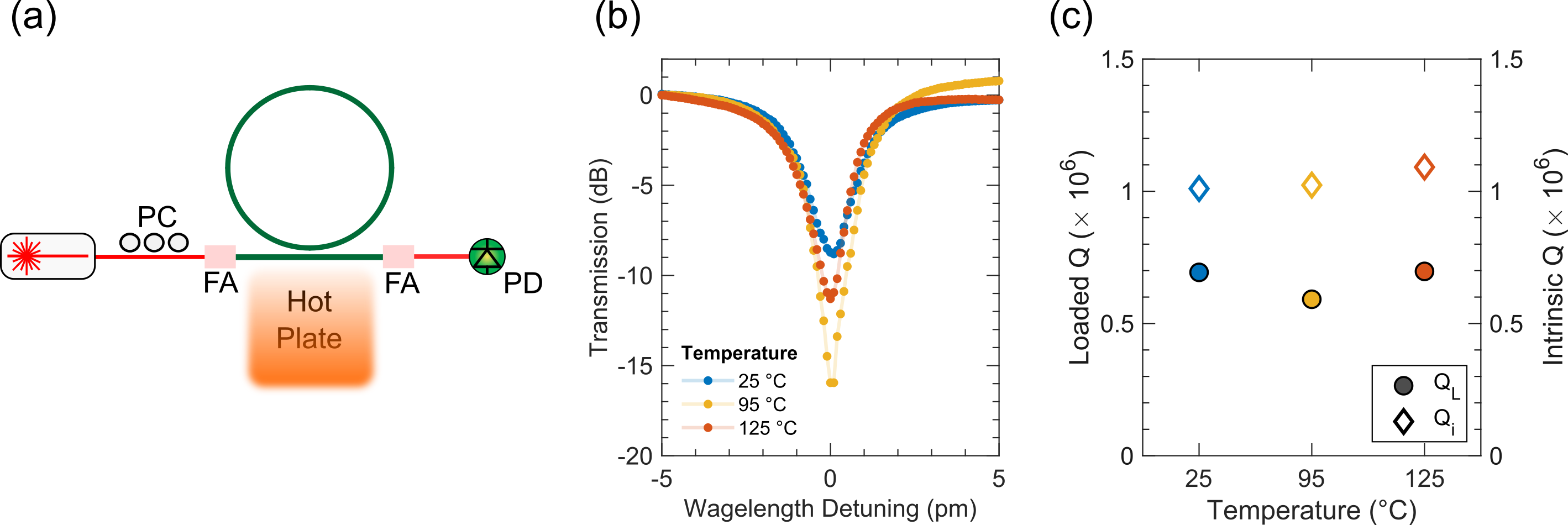}
    \caption{\textbf{High-temperature characterization of the TFLT microring resonator:} (a) Schematic of the variable-temperature measurement setup, incorporating a temperature-stabilized hot-plate, a tunable laser, polarization controller (PC), photodetector (PD), and a fiber array (FA). (b) Transmission spectra, and (c) loaded (circles) and intrinsic (diamonds) Q-factor at three different operating temperatures.}
    \label{Fig0}
\end{figure*}

\begin{figure*}[ht]
    \centering
    \includegraphics[width = 0.8\textwidth]{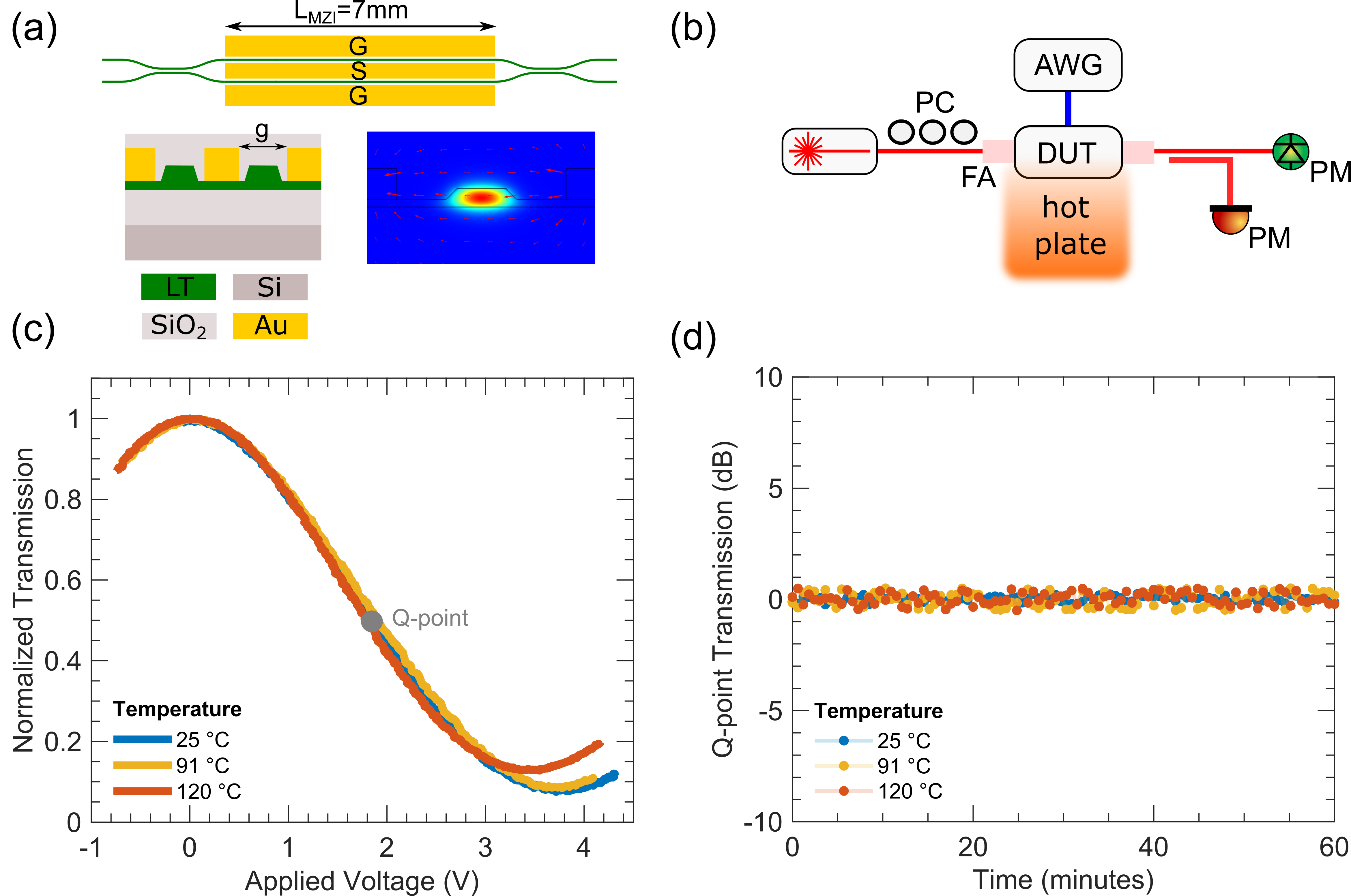}
    \caption{\textbf{TFLT Mach-Zehnder Modulator (MZM) Characterization:}(a) Top-down schematic of the traveling wave electro-optic modulator used in this experiment. The inset shows the cross-section of the MZI waveguides and the electric-field distribution of the optical and the RF mode with a gap between the signal (S) and ground (G) electrode, $\mathrm{g=\,5.5\mu m}$. (b) Schematic of the experimental setup featuring a laser source, polarization controller (PC), and the Device Under Test (DUT) mounted on a temperature-controlled hot plate. (c) Transmission as a function of applied voltage for three different operating temperatures. (d) Transmission at the Q-point over an hour across the tested temperature range.}
    \label{Fig1}
\end{figure*}

\section{Experimental results}
\begin{figure}[ht]
    \centering
    \includegraphics[width = 0.40\textwidth]{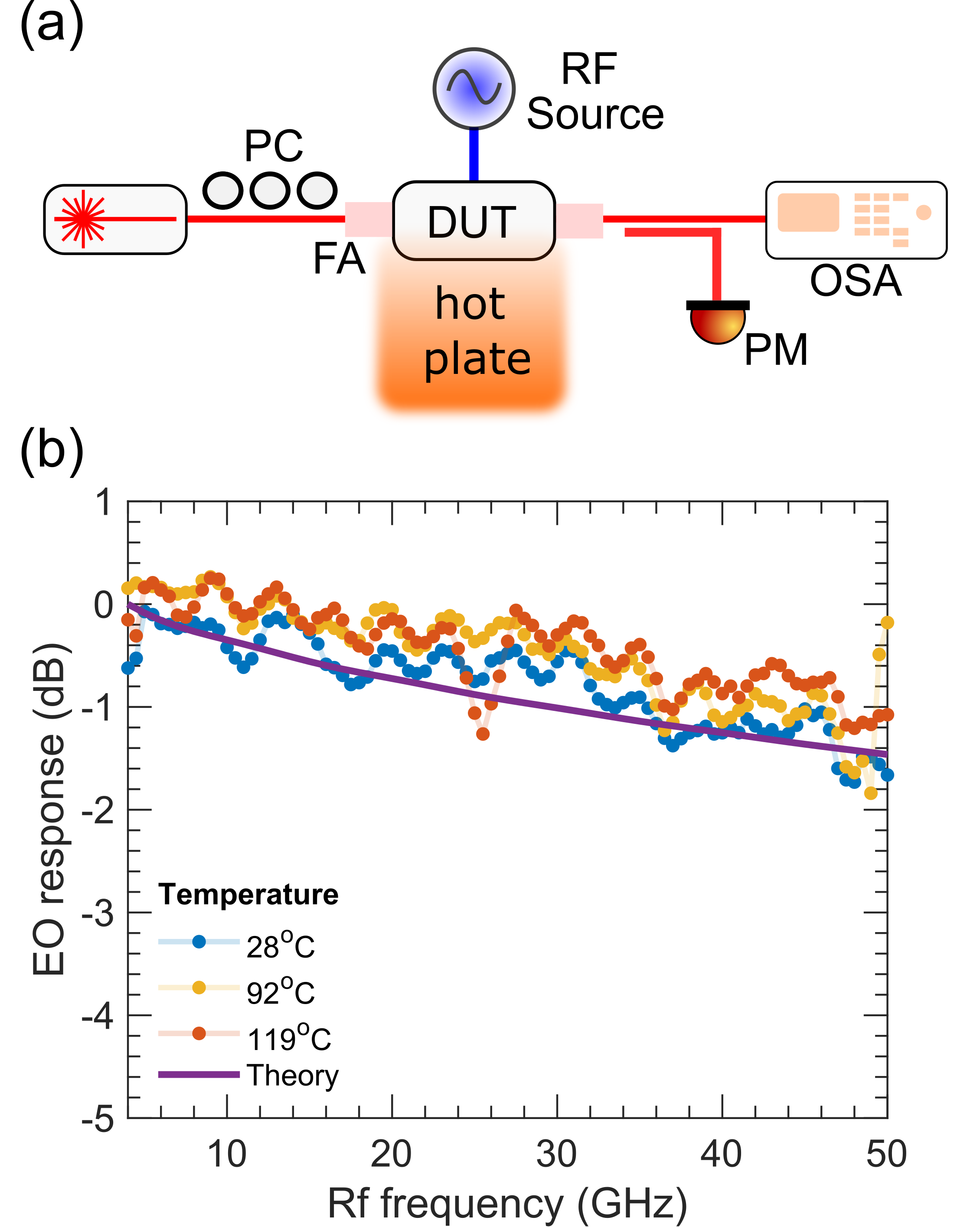}
    \caption{\textbf{TFLT Mach-Zehnder Modulator (MZM) Electro-optic Characterization:}(a) Schematic of the measurement setup. (b) Electro-optic response of the modulator as a function of RF drive frequency at three different operating temperatures.}
    \label{Fig2}
\end{figure}
We first investigate the effect of temperature on the propagation loss of the microring resonators on TFLT at telecom band. The experimental setup for characterizing the microring resonators is shown in Fig.\ref{Fig0}(a). The width and radius of the microring resonator are $\mathrm{1.8\,\mu m}$, and $\mathrm{80\,\mu m}$ respectively. Light from a tunable laser (Santec TSL-570) is coupled into the device using a fiber array (FA), with a polarization controller (PC) used to excite the fundamental TE mode. To maintain thermal stability, the DUT is mounted on a temperature-stabilized hot plate monitored by a thermocouple. The output signal is collected by a second FA and sent to a high-speed photodetector (PD) for dynamic characterization. The transmission spectra of a resonance centered at $1544.3\text{ nm}$ is shown in Fig.\ref{Fig0}(b) at three distinct temperatures: $25$°C, $95$°C, and $125$°C respectively. We extracted the loaded quality factor ($Q_L$, circles) via Lorentzian fitting and calculated the intrinsic quality factor ($Q_i$, diamonds) under the assumption of an undercoupled regime, as shown in Fig. \ref{Fig0}(c). The invariance of the $Q_i$ suggests that the propagation loss is largely unaffected by the elevated temperatures.

Fig.\ref{Fig1}(a) shows the schematic of the measurement setup used to characterize the MZM. Light from a tunable laser (Santec 570) is coupled into the device using a high numerical aperture (NA=0.41) fiber array (FA) after passing through a fiber-based polarization controller (PC). The output from the device is collected by a second fiber array (FA), split using a fiber-based power splitter, and routed to a fast photo-detector (PD) and a slow power meter (PM), respectively. An arbitrary waveform generator (AWG) provides the DC bias and RF drive signals, which are delivered to the modulator through a high-speed RF probe. The device is placed on top of a hot plate, and the hot plate temperature is monitored using a thermocouple. In Fig.\ref{Fig1}(b), we plot the normalized transmission from the device as a function of the applied voltage at three different operating temperatures, $\mathrm{25^{o}C}$, $\mathrm{91^{o}C}$, $\mathrm{120^{o}C}$, respectively. The measured $\mathrm{V_{\pi}}$ of the modulator is 3.8\,V 3.6\,V and 3.4\,V respectively. We observe a $\mathrm{\sim10\%}$ decrease in the $\mathrm{V_{\pi}}$ at elevated temperature. This is consistent with the temperature dependence of the electro-optic (EO) properties of ferroelectric materials such as LN and LT. For example, the EO effect drops $\mathrm{20-30\,\%}$ at cryogenic temperature \cite{herzog2008electro,xu2021bidirectional,Holzgrafe2020Cavity, shen2024photonic} for LN. In our case, we observe a $\mathrm{10\,\%}$ decrease of the $\mathrm{V_{\pi}}$ of the modulator at $\mathrm{120^{o}C}$ consistent with the results observed at cryogenic temperature \cite{herzog2008electro,xu2021bidirectional,shen2024photonic,Holzgrafe2020Cavity}. So, operating at a higher temperature is even beneficial for the TFLT modulators. In Fig.\ref{Fig1}(d), we show the transmission of the modulator as a function of time at three different operating temperatures when the modulator is biased at the quadrature point. From Fig.\ref{Fig1}(d), it can be observed that the modulator bias stays at the quadrature point (Q-point) without any long-term drift for over an hour. There are fluctuations around the Q-point. We suspect that the fluctuation arises mostly from hot air flow rising from the hot plate, which modifies the coupling from the fiber array to the chip. We don't observe any monotonic drift in the electro-optic bias oven one hour, showing intrinsic bias stability of the TFLT modulators even at high operating temperatures. 

\begin{figure*}[ht]
    \centering
    \includegraphics[width = 1.0\textwidth]{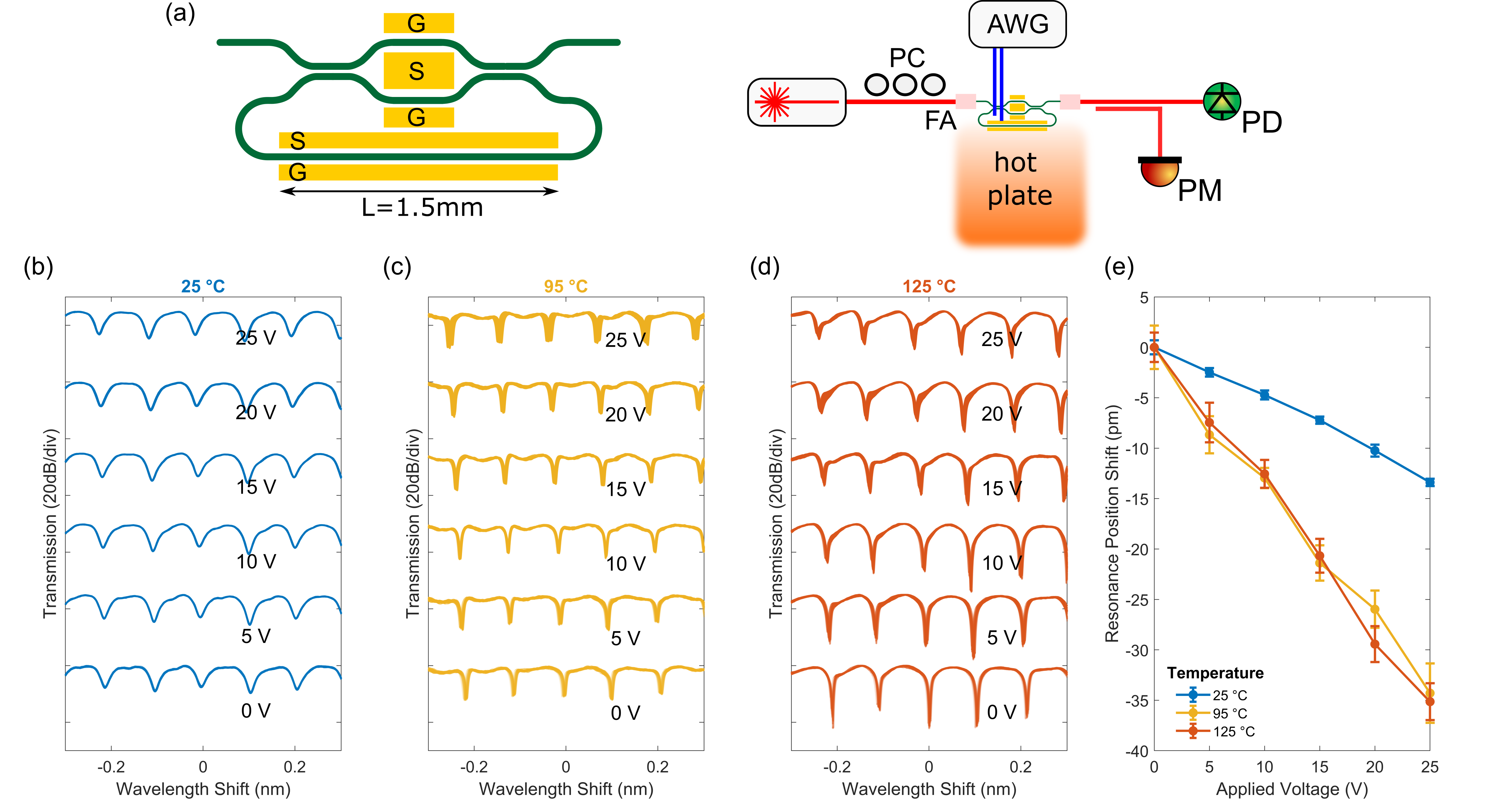}
    \caption{\textbf{Electro-optic resonance tuning of microring resonators at high temperature} (a) Schematic resonant modulator and the measurement setup, The length of the phase shifter electrode is $\mathrm{L=1.5\,mm}$ and the gap between the ground and signal electrode is $\mathrm{7\,\mu m}$. (b–d) Time stability of the transmission spectra as a function of wavelength shift across three temperatures: (a) $25^\circ$C, (b) $95^\circ$C, and (c) $125^\circ$C. The traces are vertically offset for clarity, illustrating the spectral response at applied bias voltages ranging from 0\,V to 25\,V. For each voltage, there are 20 traces, each taken one minute apart. (d) Extracted resonance wavelength shift in picometers (pm) versus the applied voltage for the three temperature conditions. Error bars represent the standard deviation across time.}
    \label{Fig3}
\end{figure*}

We measure the electro-optic (EO) response by driving the MZI modulator with an RF source and by measuring the generated optical sideband using an optical spectrum analyzer (OSA) \cite{Juneghani2022Adpr170GHz,Zhang2023LSA_TopologicalModulator,Zhang2016OL_CalibrationFreeMZM,Shi2003JLT_OSACharacterization,Purushothaman2025Optik_InterferometerFree}. The measurement setup is shown in Fig.\ref{Fig2}(a). The measurement setup is similar to the one shown in Fig.\ref{Fig1}(b) except that the AWG and the PD are replaced by a high-frequency RF source and the OSA. In Fig.\ref{Fig2}(b), we plot the electro-optic response, $\mathrm{S_{EO}}$, from 3.0\,GHz to 50\,GHz. We also plot the simulated $\mathrm{S_{EO}}$ for comparison. We expect our 3\,dB bandwidth to be beyond the 50\,GHz even assuming 1\,dB additional roll-off at low frequency. Here, our emphasis is on the temperature independence of the EO response. From Fig.\ref{Fig2}(b), it can be observed that there is no major difference in the EO response at higher frequency at different operating temperatures, showing the possibility of uncooled operation for CPO.

\section{Resonant modulator bias stability at high operating temperature}
Resonant-based modulation has been widely proposed for both long-distance communication and also for CPO-related applications \cite{Wade2020TeraPHY_IEEEMicro,Meade2019TeraPHY_OFC,Sun2020TeraPHY_VLSI,Moazeni2017RingResonatorOpticalDAC}. Although Silicon microring modulators have been widely explored, they are very sensitive to temperature fluctuations. Resonant-enhanced modulation can also be utilized in EO materials such as TFLN or TFLT \cite{xue2022breaking,sayem2026high}. For example, coupling modulators in the TFLN platform have been demonstrated with unprecedented bandwidth with a low $\mathrm{V_{\pi}}$ \cite{xue2022breaking}, but the TFLN resonator suffers drastically from the PR effect \cite{xu2021mitigating} as mentioned earlier. We have recently shown that TFLT resonators and coupling modulators are stable at high optical power and also DC-bias stable \cite{sayem2026high}. We now study the case where the resonant devices are operated at a very high temperature, similar to the traveling-wave waveguide modulator. For CPO, such compact modulators are necessary as integration density is a key parameter for CPO. We investigate the resonance tuning of such modulators at elevated temperature. In Fig.\ref{Fig3}(a), we show the schematic of the coupling electro-optic modulator \cite{sayem2026high} used in the experiment, along with the measurement setup. Fig.\ref{Fig3}(b), (c), and (d) show the transmission from the device at different operating temperatures when the phase section of the resonant modulator is biased with a DC voltage from 0\,V to 25\,V. For each voltage, we plot 20 stacked traces, where traces are recorded with a one-minute interval between each sampling. All the traces for each voltage over the measurement durations essentially overlap with each other, showing a stable bias of the resonance wavelength. In Fig.\ref{Fig3}(e), we plot the resonance frequency shift as a function of applied voltage. At room temperature, the resonator is under-coupled. At elevated temperature, the resonances become critically coupled as the coupling condition is modified by the thermo-optic effect \cite{kundu2025periodically}. This causes the EO tuning efficiency to be different at different room temperatures and at elevated temperatures. Here, we note that the change of coupling condition is greater than the microring resonator, as for the microring resonator, point coupling was used, which is significantly shorter than the coupling section for the coupling modulator. 

\section{Conclusion}
In conclusion, we show that TFLT modulators can operate at elevated temperatures without any adverse effect. High-temperature operation can even lower the $\mathrm{V_{\pi}}$ of the modulator. Most importantly, we show that resonance tuning is possible with electro-optic bias, which can eliminate all thermo-optic tuners, drastically lowering the power consumption. 

\section{Author contribution}
A.S. designed the photonic devices and developed the fabrication process flow with T.H, A.T, M.C., and R. K.. T.H, A.T, M.C., and R. K. fabricated the devices. A.S and S.U performed the measurements. A.S, S.U wrote the paper with technical feedback from M.E.

\section{Funding}
Nokia Corporation of America.

\bibliography{Reference}

\end{document}